\documentclass{PoS}

\input{psfig}
\def\farcs    {\hbox{$.\!\!^{\prime\prime}$}}
\def\kms      {\ifmmode {\rm km\,s}^{-1} \else km\,s$^{-1}$\fi}

\def\mujybm   {${\rm \mu}$Jy\,beam$^{-1}$}

\def\degr     {\hbox{$^\circ$}}

\title{Resolving the masers in M82}

\ShortTitle{Resolving the masers in M82}

\author{\speaker{Megan K. Argo}\\
        Netherlands Institute for Radio Astronomy (ASTRON), Dwingeloo, the Netherlands\\
        E-mail: \email{argo@astron.nl}}
\author{Rob J. Beswick, Tom W. B. Muxlow\\
        Jodrell Bank Centre for Astrophysics, University of Manchester, UK\\
        E-mail: \email{rbeswick@jb.man.ac.uk; twbm@jb.man.ac.uk}}
\author{Danielle Fenech\\
	Department of Physics and Astronomy, University College London, UK\\
	E-mail: \email{dmf@star.ucl.ac.uk}}
\author{Huib Jan van Langevelde\\
	Joint Institute for VLBI in Europe (JIVE), Dwingeloo, the Netherlands\\
	E-mail: \email{langevelde@jive.nl}}
\author{Melanie Gendre, Alan Pedlar\thanks{Emeritus}\\
        Jodrell Bank Centre for Astrophysics, University of Manchester, UK\\
        E-mail: \email{mgendre@jb.man.ac.uk; ap@jb.man.ac.uk}}

\abstract{Despite first being detected in the 1970s, surprisingly little is known about the OH main line maser population in the nearby starburst galaxy M82.  Sometimes referred to as `kilomasers', they have isotropic luminosities intermediate between Galactic masers and those found in more distant megamasers.  Several observations have been carried by this group over the last ten years in an attempt to get a better handle on their nature. High velocity resolution VLA observations in 2006 showed that almost all of the maser spots, distributed across the central arcminute of the galaxy, were apparently coincident with background continuum features, and a handful displayed multiple velocity components. The majority of those with velocity structure are located on a blue-shifted arc in the pv-plane, spatially located on an arc northward of the peculiar source known as B41.95+57.5. Now, new results from high spatial and spectral resolution observations with the EVN have resolved several of these masers into multiple spatial components for the first time. The maser emission is compared with known continuum sources in the galaxy, and we conclude that at least some of the maser emission is from high-gain maser action.}

\FullConference{11th European VLBI Network Symposium \& Users Meeting\\
                 9-12 October 2012\\
                 Bordeaux (France)}

\begin{document}

\section{The story so far}

The nearby galaxy M82 is one of the closest and best-studied starburst systems.  A close encounter with M81 some 220 million years ago triggered a period of rapid star formation in the core \cite{2009ApJ...701.1015K}, resulting in a high supernovae rate.  Radio studies of this galaxy are both numerous and ongoing, examining the gas and comparing the dynamics with those measured by other tracers e.g.  \cite{1984A&A...137..335W,2000MNRAS.316...33W,2002MNRAS.331..313W}, monitoring and searching for new radio supernovae and other transients \cite{2010A&A...516A..27B,2010MNRAS.404L.109M}, and investigating the nature and variability of the existing compact components \cite{2002MNRAS.334..912M,2006MNRAS.369.1221B,2010MNRAS.408..607F,2012arXiv1209.6478G}.
The galaxy is also a known source of maser emission, with OH main line emission first detected by Rieu et al \cite{1976A&A....52..467N} using the Effelsberg telescope in the 1970s.  Later observations by Weliachew et al \cite{1984A&A...137..335W} looked at the H{\sc i} and OH within the disk of M82 and suggested that the masers were not necessarily brighter than Galactic maser spots if each region contained $>$100 individual spots.  Seaquist, Frayer and Frail (1997) \cite{1997ApJ...487L.131S} later also discovered emission in the OH satellite lines.

Low spatial resolution observations of M82 were carried out with the VLA in 2002 in order to investigate the OH gas distribution in the central kiloparsec and provide a comparison with the neutral (H{\sc i}) gas dynamics where, in addition to the rotation of the disk, several shell-like features are seen in optical depth maps \cite{2002MNRAS.331..313W}.  These observations had wide ($\sim$17\,\kms) channels in order to map the absorption, resulting in dilution of the intrinsically narrow-line maser features within the broad channels.  Despite this limitation, several OH maser features were discovered in this dataset \cite{2007MNRAS.380..596A}.
These detections were followed up in 2006, again using the VLA but with much higher velocity resolution, and some new features were discovered \cite{2010MNRAS.402.2703A}.  Several features were also resolved spectrally, splitting into several velocity components and suggesting the presence of multiple maser spots within the VLA beam.
Several of the masers discovered in these two VLA datasets have apparent continuum associations, mostly with known HII regions although one is coincident with an SNR and a water maser.  But are these physical associations (so maser action as a result of star formation), or purely line-of-sight effects (so low-gain, long path length masers amplifying background continuum sources)?  With VLA-only data, it is not possible to tell.

Since the original OH observations were carried out with a view to examining the distribution of absorption across the disk, these data were plotted on a PV-diagram for direct comparison with the H{\sc i} data.  This clearly shows the main distribution which would be expected from a rotating disk, but there is also a blue-shifted arc containing the brighter masers.  This arc corresponds to a similar feature seen in H{\sc i} absorption, and the NeII emission \cite{1995ApJ...439..163A}.  The masers in this arc are spatially located in a "hole" in the 408-MHz map \cite{1997MNRAS.291..517W}, a region of absorption some 100\,pc across which may be a giant H{\sc ii} region photoionised by a cluster of early-type stars.  The 2006 VLA data clearly show that some of the masers in this region consist of multiple spectral features, so the obvious hypothesis is that these masers consist of multiple spatial components as well.
This is the question we address here.

\begin{table*}
\begin{center}
\begin{tabular}{ccccc}
\hline
Year	& Telescope	& Angular resolution	& Velocity resolution	& Notes		\\
\hline
2002	& VLA-A		& 1\farcs4		& 17\,\kms		& AP429 \cite{2007MNRAS.380..596A}	\\
2004	& EVN		& 30 mas		& 1.4\,\kms		& EB026; this paper	\\
2006	& VLA-A		& 1\farcs4		& 1.4\,\kms		& AA302 \cite{2010MNRAS.402.2703A}	\\
2012	& $e-$MERLIN	& 150 mas		& 22\,\kms		& Part of LeMMINGS \\
2012	& EVN		& 30 mas		& 1.4\,\kms		& EA051; awaiting correlation \\
\hline
\end{tabular}
\caption{\label{tab_obs}Observations of M82 carried out by this group at 1.6\,GHz.  The proposal code is noted, where appropriate.}
\end{center}
\end{table*}

Table \ref{tab_obs} details the observations of M82 at 1.6\,GHz carried out by this group in our investigations of the maser population in M82.  The distance to M82 is taken to be 3.6\,Mpc \cite{1994ApJ...427..628F} and the systemic velocity used is 225\,\kms.  Note that, while the masers are labelled according to their J2000 positions relative to 09$^{\rm h}$55$^{\rm m}$ +69\degr40' , the names of continuum features in M82 are traditionally labelled according to their B1950 positions and hence are prefixed here with "B" whenever they are referred to.

\begin{figure*}
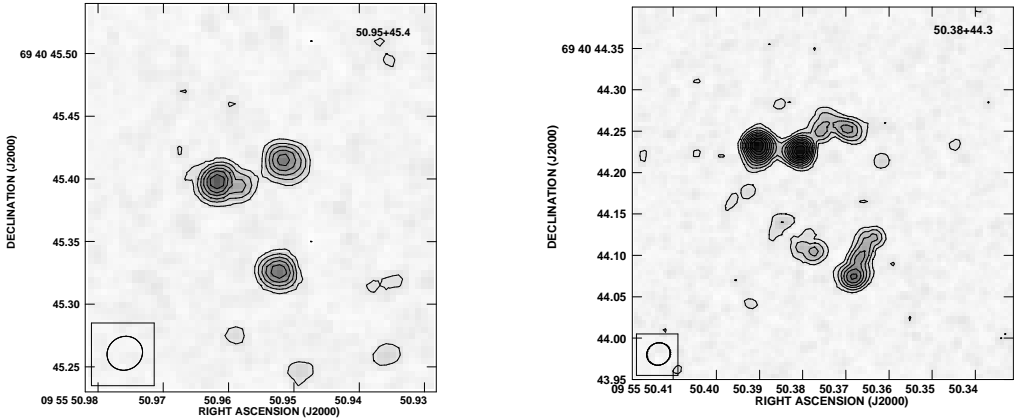

\begin{center}
	\includegraphics[width=6cm]{50.95+45.4_kntr.ps}
	\includegraphics[width=6.5cm]{50.38+44.3_kntr_90.ps}
\end{center}
\caption{\label{figure_maps}Maps of two of the EVN-detected maser regions showing their spatial extent.  The maps are from images created by collapsing each spectral line cube to a single plane using the AIPS task {\sc sqash}, taking the maximum value of each pixel, so includes emission from both OH main lines.}
\end{figure*}

\section{A closer look}

EVN observations of M82 with high angular and spectral resolution were carried out in 2004 (PI: Beswick, see Table \ref{tab_obs}).  The observation lasted 18 hours and included eight stations.  Since the size of M82 is $\sim$1 arcminute in the radio, the observations were carried out in wide-field mode; the narrow spectral channels required for the high velocity resolution allowed us to avoid bandwidth smearing, and a correlator integration time of 0.5-seconds reduced the effects of smearing due to time-averaging.  These parameters resulted in a dataset of 96-GB which was very large (for the epoch) and proved to be problematic to analyse with the resources available at the time.  Datasets of such a size are trivial by the standards of today, however, and calibration and imaging proved to be straightforward on a modern computer.

Of the thirteen known OH main line masers in M82, eight were detected in this observation.  These new maps show that most of the maser regions are still not spatially resolved, but these unresolved masers also do not show velocity sub-structure in the VLA data so this is perhaps not so surprising.  However, four of the VLA-detected maser regions, including two of the brightest masers in M82, are clearly resolved on EVN-scales.   Figure \ref{figure_maps} shows images of the two brightest masers, made from the EVN data using natural weighting.  The maps show the total line emission for each maser region, although in both cases there is considerable variation in line ratios between the individual component maser spots.  Of the detected masers, the EVN observations recover between 50 and 100 per cent of the VLA-detected flux with a beam some 40 times smaller.  Even modifying the weighting scheme to use a larger beam, we still do not recover the total VLA 2006 flux, obtained with spectral channels of about the same width, so the question is whether this is due to variability, the effects of spatial sampling, or a combination of the two.

These high-spatial resolution observations also allow us to investigate the question of whether any of the masers are physically associated with the apparently co-located continuum sources seen in the low-resolution VLA maps.  The continuum map made from the EVN data shows little continuum emission since the baselines used are sensitive only to the most compact structures.  The compact source B41.95+57.5 is obvious, however, and its flux density compares well with other high-resolution data taken a few years earlier \cite{2006MNRAS.369.1221B}.  A comparison with 5-GHz MERLIN data \cite{2008MNRAS.391.1384F} shows that in most cases there is no continuum emission at the maser location above the noise at a resolution of $\sim$60\,mas.  When comparing with the 1.7-GHz global+MERLIN data \cite{2010MNRAS.408..607F}, none of the H{\sc ii} regions are detected at all to a limit of $\sim$30\mujybm\ $-$ their structures are more diffuse than the more compact supernova remnants and the array used is not sensitive to these angular scales.  Since we see no co-located continuum emission in the EVN dataset, we are only able to put limits on the line to continuum ratios for each component.

One source which shows some coincidence with continuum emission from a higher resolution dataset is 50.95+45.4 which appears to be associated with B42.41+59.2, an H{\sc ii} region \cite{2002MNRAS.334..912M}.  This source (see left panel in Fig.~\ref{figure_maps}) is composed of several OH emission components which are spread over $\sim$11\,\kms\ in velocity and spatially align roughly with a ring of emission seen in the 5-GHz MERLIN map \cite{2008MNRAS.391.1384F}, although the individual maser spots appear to avoid the peaks of the continuum emission.  This could be due to registration error between the two datasets, so this effect needs to be carefully investigated.

The brightest OH maser is 50.38+44.3, also spatially located within the 408-MHz ring of absorption \cite{1997MNRAS.291..517W} and on the blue-shifted arc in the p-v diagram.  As for 50.95+45.4, the velocities of the components are spread over $\sim$10\,\kms, a smaller distribution than that seen for either maser in the lower spatial resolution VLA 2006 data.  Most of the components are only seen in the 1667-MHz line with only a couple of weak components visible at 1665-MHz, recovering less than 50\% of the VLA2006 flux.  This region is coincident with the continuum source B41.64+57.9, another probable H{\sc ii} region with little emission on MERLIN-scales at 5\,GHz \cite{2008MNRAS.391.1384F}.  As above, the maser emission again appears to avoid the peaks of 5-GHz emission, and the same caveat applies.  A comparison with 15-GHz VLA+PT data \cite{2002MNRAS.334..912M} suggests that this source has a very weak predicted L-band flux on these scales.  With the number of components seen in this region, it is interesting to note that there appears to be a velocity gradient across the map with the northern ridge-like feature blue-shifted compared to the southern complex.

Observations of OH masers in Arp220 show some similarities to the masers in M82.  Much of the total flux density of the maser emission measured at low-resolution was resolved out when observed at higher angular resolution, leading to the conclusion that this megamaser system is composed of both diffuse (low-gain, unsaturated, IR-pumped) emission {\bf and} compact (amplified, saturated, collisionally-pumped) emission \cite{1998ApJ...493L..13L}.  The same authors also found no continuum at the maser locations, giving lower limits on the amplification factors of up to 800.  The line ratios are also very high, with no 1665-MHz emission detected at mas resolutions.
The compact masers in Arp220 show significant velocity width, wider than the intrinsic $\sim$1\,\kms, but are spatially unresolved; the M82 masers here which show similar velocity widths are precisely the ones which $are$ resolved into multiple spatial components, suggesting that the same may be true for the Arp220 masers if we had the resolution to probe such scales at the greater distance of Arp220 (their observations had a resolution of 3.1 x 8.0 mas, or 1.0 x 2.5 pc for Arp220 at an assumed distance of 76 Mpc).
Unlike in Arp220, there is compact maser emission in M82 on mas scales in the 1665\,MHz line, in Arp220 the only 1665\,MHz emission is diffuse and hence resolved out on VLBI scales.

\section{To be continued}

The maser population in M82 is interesting, and these new EVN results are adding important information to the overall picture.
The question of association with continuum sources can be investigated by a careful comparison of these images with continuum datasets using similar resolution.  Limits on the line-to-continuum ratios can be measured directly from the EVN data, as well as the ratio between the 1665- and 1667-MHz lines for each individual maser spot.
First indications are that compact OH maser emission is seen offset from the continuum, demonstrating that high-gain maser action may be responsible for a part of the maser emission in M82.

Observations at low-velocity resolution and intermediate angular resolution have been carried out recently with e-MERLIN as part of the LeMMINGS Legacy survey, and further high-velocity resolution observations have been requested to match the velocity resolution of the EVN observations.  These observations should enable a search for a more diffuse maser component such as that observed in Arp220.

One question it has so far been impossible to answer is that of variability.  So far, each observation has used dramatically different parameters in either spectral or angular resolution, making both a comparison of brightnesses between epochs and a search for morphological changes impossible.  To remedy this situation, a second EVN epoch was observed in the most recent EVN session; eight years after the first, but with the same parameters.

\bibliographystyle{JHEP-2}
\bibliography{refs}

\end{document}